\documentclass[%
 reprint,
 amsmath,amssymb,
 aps,
]{revtex4-1}
\usepackage{graphicx}
\usepackage{color}
\usepackage{bm}
\usepackage{epsfig}

\def \a{\alpha}
\def \b{\beta}

\def \g{\gamma}
\def \d{\delta}
\def \s{\sqrt}
\def \be{\begin{equation}}
\def \ee{\end{equation}}
\def \ben{\begin{eqnarray}}
\def \een{\end{eqnarray}}

\def \D{\Delta}

\def \t{\theta}
\def \P{\Phi}
\def \i{\eta}
\def \m{\mu}
\def \x{\xi}
\def \c{\chi}
\def \p {\psi}
\def \n {\nu}

\begin{document}

\title{Time-like geodesic structure for the K-essence Emergent Barriola-Vilenkin type spacetime}

\author{Bivash Majumder}
\altaffiliation{bivashmajumder@gmail.com}
\affiliation{Department of Mathematics, Prabhat Kumar College, Contai, Purba Medinipur-721404, India}

\author{Goutam Manna }
\altaffiliation{goutammanna.pkc@gmail.com}
\affiliation{Department of Physics, Prabhat Kumar College, Contai, Purba Medinipur-721404, India}

\author{Ashoke Das }
\altaffiliation{ashoke.avik@gmail.com}
\affiliation{Department of Mathematics, Raiganj University, Raiganj, West Bengal-733134, India}

\begin{abstract}
For a particular type of {\bf k-}essence scalar field, the {\bf k-}essence emergent gravity metric is exactly mapped on to the Barriola-Vilenkin (BV) type metric for Schwarzschild background established by Gangopadhyay and Manna.
Based on the S. Chandrasekhar, we report the exciting features of the time-like geodesic structure in the presence of dark energy in an emergent gravity scenario for this Barriola-Vilenkin type metric. We trace the different kinds of trajectories for time-like
geodesic in the presence of dark energy for the {\bf k-}essence emergent Barriola-Vilenkin spacetime, which is same as the
Schwarzchild spacetime in view of the basic orientation, but the allowed ranges of the aphelion and
perihelion distances are much more different. The bound and unbound orbits are plotted for a fixed
value of the dark energy density.
\end{abstract}


\pacs{..................}

\maketitle

\section{Introduction}
The time-like geodesic structures of the Schwarzschild spacetime 
have been discussed in detail by S. Chandrasekhar in his book \cite{chandra}, ({\it chapter-3, sections 19}).
He has also discussed the orbital structures of the bound and unbound orbits with figures. Also
in \cite{cruz}, authors have studied the geodesic structures for the Schwarzschild anti-de Sitter spacetime. They evaluated radial and non-radial trajectories for time-like and null geodesics. They also have shown that the geodesic structures of this black hole presents new types of motion which is not allowed by the Schwarzschild spacetime. 
The geodesic structure of the Schwarzschild spacetime also discussed in \cite{berti}. The Jacobi metric for time-like geodesics in
static spacetimes have been discussed in \cite{gibbons}. They have shown that the free motion of massive particles moving in static spacetimes is given by the geodesics of an energy-dependent Riemannian metric on
the spatial sections analogous to Jacobi's metric in classical dynamics. In the
massless limit Jacobi's metric coincides with the energy independent Fermat
or optical metric. They have also described the properies of the Jacobi
metric for massive particles moving outside the horizon of a Schwarschild
black hole. In \cite{chanda}, they obtained the Jacobi metric for various stationary metrics and formulated the Jacobi-Maupertuis
metric for time-dependent metrics by including the Eisenhart-Duval lift \cite{eisenhart,duval}.

Barriola and Vilenkin
established the solution for Einstein equation outside the monopole core \cite{barriola}, where a global monopole falls into a Schwarzschild black hole the resulting black hole carries the global monopole charge.
In \cite{gm1}, based on the Dirac-Born-Infeld (DBI) model \cite{born1,born2,born3} of the  {\bf k-}essence theory \cite{babi1,babi2,babi3,babi4,babi5}, the authors have shown that for a particular configuration of {\bf k-}essence scalar field $(\phi)$ the emergent gravity metric $\bar G_{\mu\nu}$ is conformally equivalent to the Barriola-Vilenkin (BV) metric for Schwarzschild background where global monopole charge replaced by the constant kinetic energy ($\dot\phi^{2}=K$) of the {\bf k-}essence scalar field. The emergent gravity metric ($\bar G_{\mu\nu}$) is not conformally equivalent to the gravitational metric ($g_{\mu\nu}$). The Lagrangian 
contains non-canonical kinetic terms for the $k-$essence scalar fields. 
The general form of the lagrangian for $k-$essence model is: $L=-V(\phi)F(X)$ where $X=\frac{1}{2}g^{\mu\nu}\nabla_{\mu}\phi\nabla_{\nu}\phi$ and it does not 
depend explicitly on $\phi$ to start with \cite{gm1,babi1,babi2,babi3,babi4,babi5,scherrer1,scherrer2}. 
The difference between the {\bf k-}essence theory with non-canonical kinetic terms and the
relativistic field theories with canonical kinetic terms is that
the nontrivial dynamical solutions of the {\bf k-}essence equation of motion not only spontaneously break Lorentz invariance but also change the metric due to the perturbations around these solutions. So the perturbations propagate in the {\it emergent} or analogue curved spacetime \cite{babi1,babi2,babi3,babi4,babi5} with the metric different from the gravitational one. 

In this work, we investigate the time-like geodesic structures for the {\bf k-}essence emergent Barriola-Vilenkin (BV) type spacetime in the presence of dark energy based on the Ref. \cite{chandra} not in the context of Jacobi metric \cite{gibbons,chanda}.

The paper is organized as follows:
firstly, we have briefly described the {\bf k-}essence theory   and emergent gravity where the metric $\bar G_{\mu\nu}$ contains the 
dark energy field $\phi$ and it should satisfy the emergent gravity equations of motion in section-2. 
Again, for  $\bar G_{\mu\nu}$ to be a black hole metric, it has to satisfy the Einstein field equations.
In section 3, we describe the geodesics of the BV type emergent spacetime. In section-4, we detail discuss the structure of the time-like geodesic of the Barriola-Vilenkin type emergent spacetime including bound and unbound orbits.
The last section is the conclusion of our work.

\section{Review  of {\bf k}-essence theory and Emergent Gravity}

In this section, we present a short review of the {\bf k-}essence theory and construction of the effective emergent metric. The {\bf k}-essence scalar field $\phi$ minimally coupled to the background spacetime metric $g_{\mu\nu}$ has action \cite{babi1}-\cite{babi5}
\ben
S_{k}[\phi,g_{\mu\nu}]= \int d^{4}x {\sqrt -g} L(X,\phi)
\label{1}
\een
where $X={1\over 2}g^{\mu\nu}\nabla_{\mu}\phi\nabla_{\nu}\phi$.
The energy-momentum tensor is
\ben
T_{\mu\nu}\equiv {2\over \sqrt {-g}}{\delta S_{k}\over \delta g^{\mu\nu}}= L_{X}\nabla_{\mu}\phi\nabla_{\nu}\phi - g_{\mu\nu}L
\label{2}
\een
$L_{\mathrm X}= {dL\over dX},~~ L_{\mathrm XX}= {d^{2}L\over dX^{2}},
~~L_{\mathrm\phi}={dL\over d\phi}$ and  
$\nabla_{\mu}$ is the covariant derivative defined with respect to the gravitational metric $g_{\mu\nu}$.
The scalar field equation of motion is
\ben
-{1\over \sqrt {-g}}{\delta S_{k}\over \delta \phi}= G^{\mu\nu}\nabla_{\mu}\nabla_{\nu}\phi +2XL_{X\phi}-L_{\phi}=0
\label{3}
\een
where  
\ben
 G^{\mu\nu}\equiv \frac{c_{s}}{L_{X}^{2}}[L_{X} g^{\mu\nu} + L_{XX} \nabla ^{\mu}\phi\nabla^{\nu}\phi]
\label{4}
\een
and $1+ {2X  L_{XX}\over L_{X}} > 0$ with
$c_s^{2}(X,\phi)\equiv{(1+2X{L_{XX}\over L_{X}})^{-1}}$. The inverse metric of $G^{\mu\nu}$ is   
\ben G_{\mu\nu}={L_{X}\over c_{s}}[g_{\mu\nu}-{c_{s}^{2}}{L_{XX}\over L_{X}}\nabla_{\mu}\phi\nabla_{\nu}\phi] .
\label{5}
\een
Making a conformal transformation \cite{gm1,gm2,gm3} $\bar G_{\mu\nu}\equiv {c_{s}\over L_{X}}G_{\mu\nu}$ gives
\ben \bar G_{\mu\nu}
={g_{\mu\nu}-{{L_{XX}}\over {L_{X}+2XL_{XX}}}\nabla_{\mu}\phi\nabla_{\nu}\phi}
\label{6}
\een	
Here $L_{X}\neq 0$ for $c_{s}^{2}$ to be positive definite and only then equations (\ref{1})-(\ref{4}) will be physically meaningful.

Generally, the emergent gravity metric $G_{\mu\nu}$ is not conformally equivalent to the gravitational metric $g_{\mu\nu}$. So $\phi$ has properties different from canonical scalar fields, with the local causal structure also different from those defined with $g_{\mu\nu}$. Further, if $L$ is not an explicit function of $\phi$ then the equation of motion (\ref{3}) reduces to;
\ben
-{1\over \sqrt {-g}}{\delta S_{k}\over \delta \phi}
= \bar G^{\mu\nu}\nabla_{\mu}\nabla_{\nu}\phi=0
\label{7}
\een
Now we take a particular type of the  DBI Lagrangian \cite{gm1,gm2,gm3},\cite{born1}-\cite{born3} as 

\ben
L(X,\phi)= 1-V(\phi)\sqrt{1-2X}
\label{8}
\een
with $V(\phi)=V=constant$~~and~~$kinetic ~ energy ~ of~\phi>>V$ i.e.$(\dot\phi)^{2}>>V$. This is typical for the {\bf k}-essence fields where the kinetic energy dominates over the potential energy. Then $c_{s}^{2}(X,\phi)=1-2X$. For scalar fields $\nabla_{\mu}\phi=\partial_{\mu}\phi$. Thus (\ref{6}) becomes
\ben
\bar G_{\mu\nu}= g_{\mu\nu} - \partial _{\mu}\phi\partial_{\nu}\phi
\label{9}
\een

\section{Geodesics for the BV type emergent spacetime}
In \cite{gm1}, the authors have established that for the gravitational metric $g_{\mu\nu}$ is to be Schwarzschild, the emergent metric $\bar G_{\mu\nu}$ (\ref{9}) is exactly mapped on to the Barriola-Vilenkin (BV) type metric for a particular type of the {\bf k-}essence scalar field with the global monopole charge is replaced by the constant kinetic energy of the scalar field. The {\bf k-}essence emergent BV type  metric is \cite{gm1} 
\ben
ds^{2} =(1-{2GM\over r}-K)dt^{2}
-{1\over(1-{2GM\over r}-K)}dr^{2}\nonumber\\-r^{2}d\theta^{2}-r^{2}sin^{2}\theta d\Phi^{2}\nonumber\\
 = (\b-{2GM\over r})dt^{2}-\frac{dr^{2}}{(\b-{2GM\over r})}-r^{2}d\theta^{2}-r^{2}sin^{2}\theta d\Phi^{2}\nonumber\\
\label{eq:1}
\een
where $\b=(1-K)$ and $K$ is the constant kinetic energy of the {\bf k-}essence scalar field i.e., the dark energy density in unit of critical density \cite{gm1,gm2,gm3} which have values $0<K<1$ and the solution of this {\bf k-}essence scalar field \cite{gm1} is $\phi(r,t)=\phi_{1}(r)+\phi_{2}(t)
=\sqrt{K}[r+2GM~ln(r-2GM)]+\sqrt{K}t$. Here mention that though in this form of {\bf k-}essence scalar field violets the Lorentz invariance but the {\bf k-}essence theory admits this violation. In addition, it should be noted that $K$ has always in the range between $0<K<1$ since if $K=(\dot{\phi_{2}})^{2}=0$ then the {\bf k-}essence theory is meaningless and  $K=1$, the metric (10) does not have a Newtonian limit  \cite{schutz} and $K>1$ is not possible as the total energy density cannot
exceed unity ($\Omega_{matter} + \Omega_{radiation} + \Omega_{darkenergy} = 1$) also if $K>1$ then signature of the above metric (10) is ill-defined.

The geodesics equation \cite{chandra,cruz,berti} in the Barriola-Vilenkin type emergent spacetime (\ref{eq:1}) can be derived from the Lagrangian

\ben
2\mathcal{L} =(\b-{2GM\over r})\dot t^{2}
-{1\over(\b-{2GM\over r})}\dot r^{2}
-r^{2}\dot {\t}^{2}-(r^{2}sin^{2}\t) \dot {\P}^{2}\nonumber\\
\label{eq:2}
\een
where
 $\dot t = \frac{dt}{d\tau},~\dot r = \frac{dr}{d\tau},~
\dot \t = \frac{d\t}{d\tau},~\dot \P = \frac{d\P}{d\tau}$, $\tau $ is to be identified with the proper time.
The momenta associated with this Lagrangian are
$ p_{t}=\frac{\delta \mathcal{L}}{\delta \dot{t}}=(\b-{2GM\over r})\dot t,$
$p_{r}=-\frac{\delta \mathcal{L}}{\delta \dot{r}}={1\over(\b-{2GM\over r})}\dot r,$
$p_{\t}=-\frac{\delta \mathcal{L}}{\delta \dot{\t}}=r^{2}\dot{\t},$
$
p_{\Phi}=-\frac{\delta \mathcal{L}}{\delta \dot{\Phi}}=r^{2}sin^{2}\t\dot{\P}.
$ 

The Hamiltonian is
\ben
 \mathcal{H}=p_{\m}\dot x^{\m}-\mathcal{L}=p_{t}\dot{t}-(p_{r}\dot{r}+p_{\t}\dot\t+p_{\P}\dot{\P})-\mathcal{L}=\mathcal{L}\nonumber\\
\label{eq:3}
\een
The equality of Hamiltonian and Lagrangian of above equation implies that the Lagrangian (\ref{eq:2}) is purely kinetic. So there is no potential energy contribution in the problem which is also agree with the {\bf k-}essence theory since in this theory, the contribution of the kinetic energy part is dominated over the potential energy i.e., $K.E.>>P.E.$. The constancy of the Hamiltonian and Lagrangian implies that $\mathcal{H}=\mathcal{L}=constant$.

By rescaling the affine parameter $\tau$, we shall consider $2\mathcal{L}=+1$ for time-like geodesic and $2\mathcal{L}=0$ for null geodesics. Here we shall not consider null geodesics and space-like geodesics.

As a consequence of the static, spherically symmetric nature of the metric, the Lagrangian (\ref{eq:2}) does not depend on $t$ and $\Phi$.
Thus the equations of motion are,
$\dot{p_{t}}=0,~\dot{p_{\P}}=0$, this implies 
\ben p_{t}=(\b-{2GM\over r})\dot t=constant=E(say)
\label{eq:4}
\een
and
\ben p_{\P}=r^{2}sin^{2}\t\dot{\P}=constant
\label{eq:5}
\een
again from the equation of motion, 
$$ \dot{p_{\t}}=-\frac{\d L}{\d \t}
$$
\ben
\Rightarrow\frac{d}{d\tau}(r^{2}\dot{\t})=(r^{2}\sin\t\cos\t)\dot{\P}^{2}
\label{eq:6}
\een
Our object and metric are spherically symmetric we can simplify things by only considering motion
in the equatorial plane $ \t=\frac{\pi}{2}$ and $ \dot{\t}=0$. Then equation (\ref{eq:5}) gives 
\ben p_{\P}=r^{2}\dot{\P}=constant=L~(say)
\label{eq:7}
\een
where $L$ denotes the angular momentum about an axis normal to the invariant plane.
By the equation (\ref{eq:4}) and (\ref{eq:7}), we have $\dot{t}=\frac{E}{\b-\frac{2M}{r}} $ and $\dot{\P}=\frac{L}{r^{2}}$. Substituting these values in (\ref{eq:2}), the Lagrangian can be rewritten as
\ben
2\mathcal{L}=\frac{E^{2}}{\b-\frac{2M}{r}}-\frac{\dot{r}^{2}}{\b-\frac{2M}{r}}-\frac{L}{r^{2}}
\label{eq:8}
\een

\section{Time-like Geodesic for the BV type emergent spacetime}
 For time-like geodesics ($2\mathcal{L}=+1$), equation (\ref{eq:8}) can be written as 
\ben 
\dot{r}^{2}=E^{2}-\left(1+\frac{L^{2}}{r^{2}}\right)\left(\b-\frac{2M}{r}\right)
\label{eq:9}
\een
and 
\ben
\frac{d\P}{d\tau}=\frac{L}{r^{2}}
\label{eq:10}
\een
Substituting (\ref{eq:10}) into (\ref{eq:9}), we obtain the equation
\ben
\left(\frac{dr}{d\P}\right)=(E^{2}-\b)\frac{r^{4}}{L^{2}}+\frac{2M}{L^{2}}r^{3}-\b r^{2}+2Mr
\label{eq:11}
\een
Now we consider
\ben 
u=\frac{1}{r}
\label{eq:12}
\een
Then the equation (\ref{eq:11}) transformed to
\ben
\left(\frac{du}{d\P}\right)^{2}=2M u^{3}-\b u^{2}+\frac{2M}{L^{2}}u-\frac{\b-E^{2}}{L^{2}}
\label{eq:13}
\een
implies
\ben
\left(\frac{du}{d\P}\right)^{2}=\b\left[2 \bar{M} u^{3}-u^{2}+\frac{2\bar{M}}{L^{2}}u-\frac{1-\bar{E}^{2}}{L^{2}}\right]
\label{eq:14}
\een
where \textbf{ $\bar{ M}=\frac{M}{\b}$ and  
{\it $\bar{E}=\frac{E}{\s{\b}}$}}.
From the equations (\ref{eq:10}) and (\ref{eq:12})
\ben
\frac{d\tau}{d\Phi} =\frac{1}{Lu^{2}}
\label{eq:15}
\een
with help of the above equation and (\ref{eq:4}), we can write
\ben 
\frac{dt}{d\P}=\frac{\bar{E}}{\s\b (1-2\bar{M}u)}.\frac{1}{Lu^{2}}
\label{eq:16}
\een
 The geometry of the geodesics in the invariant plane shall be obtained by solving the basic equation (\ref{eq:14}) and the equations (\ref{eq:15}) and (\ref{eq:16}).

\subsection{\textbf{The radial Geodesics}}
 The radial geodesics correspond to the  motion of the particles without angular momentum $(L=0)$ which start from rest at some finite distance $r=r_{a}$ and fall towards the centre. The equations (\ref{eq:4}) and (\ref{eq:9}) transformed to 
 \ben
 \frac{dt}{d\tau}=\frac{\bar{E}}{\s\b (1-2\bar{M}u)}
 \label{eq:17}
 \een
 and 
 \ben
 \left(\frac{dr}{d\tau}\right)^{2}=\b \left[2\bar{M}u-(1-\bar{E}^{2})\right]
 \label{eq:18}
\een
 
 Clearly $\dot{r}=0$ at $r=r_{a}$. Therefore from the equation (\ref{eq:18})
\ben  
  r_{a}=\frac{2\bar{M}}{1-\bar{E}^2}
\label{eq:19}
\een
 Let us take the substitution 
\ben  
  r=\frac{\bar{M}}{1-\bar{E}^2}(1+\cos\i)=r_{a} \cos^{2}\frac{\i}{2}
\label{eq:20}
\een
 Therefore, 
 $\i=0$ when $r=r_{a}$; $\i=\pi$ at the singularity $(r=0)$ and $\i=\i_{H}=2 \sin^{-1}(\bar{E})$ when $r$ crosses the horizon $r=2\bar{M}=\frac{2M}{\b}$.
 
 So the equations (\ref{eq:17}) and (\ref{eq:18}) becomes
 \ben
 \frac{dt}{d\tau}=\frac{\bar{E}~cos^{2}(\i/2)}{\s{\b}[cos^{2}(\i/2)-cos^{2}(\i_{H}/2)]}
 \label{eq:21}
 \een
 and
 \ben
 \left(\frac{dr}{d\tau}\right)^{2}=\b(1-\bar{E}^{2})tan^{2}(\i/2)
 \label{eq:22}
 \een

Using the equations (\ref{eq:19}), (\ref{eq:20}) and (\ref{eq:22}) and considering infalling particles we get
 \ben
 \frac{d\tau}{d\i}=\frac{1}{\s\b}\left(\frac{r_{a}^{3}}{8\bar{M}}\right)^{\frac{1}{2}}(1+\cos\i)
 \label{eq:23}
 \een 
 Therefore,
 \ben
\tau=\frac{1}{\s\b}\left(\frac{r_{a}^{3}}{8\bar{M}}\right)^{\frac{1}{2}}(\i+\sin\i)
\label{eq:24}
\een 
 Here, we assumed that $\tau=0$ at $\i=0$ (i.e., at $r=r_{a}$). From previous equation (\ref{eq:24}), we can say that the particles crosses the horizon $(r=r_{a})$ and arrives  at the singularity $(r=0)$ at the finite proper times,
 \ben
 \tau_{H}=\frac{1}{\s\b}\left(\frac{r_{a}^{3}}{8\bar{M}}\right)^{\frac{1}{2}}(\i_{H}+\sin\i_{H})
\label{eq:25}
\een
and
\ben
\tau_{0}=\frac{\pi}{\s\b}\left(\frac{r_{a}^{3}}{8\bar{M}}\right)^{\frac{1}{2}}.
\label{eq:26} 
\een 
 Again from the equations (\ref{eq:21}) and (\ref{eq:23}) and integrating we get

 \ben
t_{BV}^{k}=\frac{\bar{E}}{\b}\left(\frac{r_{a}^{3}}{2\bar{M}}\right)^{\frac{1}{2}}\left[\frac{1}{2}(\i+\sin\i)+(1-\bar{E}^{2})\i\right]\nonumber\\+\frac{2\bar{M}}{\b}~ln\left(\frac{\tan\frac{\i_{H}}{2}+\tan\frac{\i}{2}}{\tan\frac{\i_{H}}{2}-\tan\frac{\i}{2}}\right)\nonumber\\
=\frac{EM}{\b(\b-E^{2})^{3/2}}(\i +\sin\i)+\frac{2EM}{\b^{2}(\b-E^{2})^{1/2}}~\i \nonumber\\+\frac{2M}{\b^{2}}ln\left(\frac{\frac{E}{\sqrt{\b-E^{2}}}+\tan\frac{\i}{2}}{\frac{E}{\sqrt{\b-E^{2}}}-\tan\frac{\i}{2}}\right)\nonumber\\
\label{eq:27}
\een
 This shows that $t\rightarrow\infty$ as $\i\rightarrow \i_{H}-0 $ i.e., a particle will take an infinite time to reach the horizon with respect to an observer stationed at infinity, even though the particle will cross the horizon in a finite time by its proper time and again it will take finite proper time to reach the singularity.
 
The usual time in the radial geodesic for Schwarzschild space time is \cite{chandra}
 \ben
t_{Sch}=E\left(\frac{r_{a}^{3}}{2M}\right)^{\frac{1}{2}}\left[\frac{1}{2}(\i+\sin\i)+(1-E^{2})\i\right]\nonumber\\+2M~ln\left(\frac{\tan\frac{\i_{H}}{2}+\tan\frac{\i}{2}}{\tan\frac{\i_{H}}{2}-\tan\frac{\i}{2}}\right)\nonumber\\
=\frac{EM}{(1-E^{2})^{3/2}}(\i +\sin\i)+\frac{2EM}{(1-E^{2})^{1/2}}~\i \nonumber\\+2M~ln\left(\frac{\frac{E}{\sqrt{1-E^{2}}}+\tan\frac{\i}{2}}{\frac{E}{\sqrt{1-E^{2}}}-\tan\frac{\i}{2}}\right)\nonumber\\
\label{eq:28}
\een
 where $ r=\frac{M}{1-E^2}(1+\cos\i)=r_{a} \cos^{2}\frac{\i}{2}$, $\i_{H}=2 \sin^{-1}(E)$ and $r_{a}=\frac{2M}{1-E^2}$.
 {\it The above two equations (\ref{eq:27}) and (\ref{eq:28}) shows that $t_{BV}^{k}>t_{Sch}$ since $K$ has values ($0<K<1$), so $\b~(=1-K)$ also the same range as $0<\b<1$.
Also note that as $K\rightarrow 1$ i.e., $\b \rightarrow 0$, $t^{k}_{BV}$ is undefined since $t^{k}_{BV}$ contain the terms $(\b-E^{2})^{3/2}$ and $(\b-E^{2})^{1/2}$  and these terms are undefined when $\b \rightarrow 0$.}

\subsection{\textbf{Bound Orbits $(\bar{E}^{2}<1)$}}  

For $\bar {E}^{2}<1$, the governing equation \cite{chandra} (\ref{eq:14})   
\ben
f(u)=\b 2\bar{M} \left(u^{3}-\frac{1}{2\bar{M}} u^{2} +\frac{1}{L^{2}}u-\frac{1-\bar{E}^{2}}{2\bar{M} L^{2}}\right)
\label{eq:29}
\een
where 
\ben
f(u)=\left(\frac{du}{d\P}\right)^{2}.
\label{eq:30}
\een
gives no negative roots, the only possibilities are positive roots and complex
roots. So in the invariant plane, $r$ remains bounded. Therefore, the orbits are
bounded in the invariant plane.

Let the roots of the cubic equation $f(u)=0$ are $u_{1},u_{2}$ and $u_{3}$, then we have 
\ben
  u_{1}+u_{2}+u_{3}=\frac{1}{2\bar{M}}
  \label{eq:31}
\een
\ben  
 u_{1}u_{2}+u_{2}u_{3}+u_{3}u_{1}=\frac{1}{L^{2}}
  \label{eq:32}
\een
\ben 
 u_{1}u_{2}u_{3}=\frac{1-\bar{E}^{2}}{2\bar{M} L^{2}}
   \label{eq:33}
\een
We have, when $u=0$, then $f(u)<0$ and when $u\to \pm\infty$ then $f(u)\to \pm\infty$. Further by our assumption $(\bar {E}^{2}<1)$, the equation $f(u)=0$ have no negative roots and it must have at least one positive roots. For every pair of values $\bar{E}$ and $L$, this leads to five different cases as discussed in the following cases

 Case $(\a)$: All the three roots are real and positive. Here we shall take $0<u_{1}<u_{2}<u_{3}$. Therefore $f(u)<0$ when $u<u_{1}$; $f(u)>0$ when $u_{1}<u<u_{2}$; $f(u)<0$ when $u_{2}<u<u_{3}$ and $f(u)>0$ when $u>u_{3}$. So there are two kinds of orbits which can be traced. One oscillates between the values $u_{1}^{-1}$ and $u_{2}^{-1}$, which will be called the orbits of first kind and the other one is starting from the distane $u_{3}^{-1}$ and plunges to the singularity $(r=0)$ which will be called the orbits of second kind.

Case $(\b)$: All the  roots are real and positive and two of the three roots are equal. Here we shall take $0<u_{1}=u_{2}<u_{3}$. Therefore $f(u)<0$ when $u<u_{1}=u_{2}$ and $u_{1}=u_{2}<u<u_{3}$ and $f(u)>0$ when $u>u_{3}$. Again there are two kinds of orbits which can be traced. One, the orbits of first kind which is a stable circular orbit with radius $u_{1}^{-1}$. The other one is the orbits of the second kind which starts from the distance $u_{3}^{-1}$ and plunges to singularity.

Case $(\g)$: All the  roots are real and positive and two of them are equal. Here we shall take $0<u_{1}<u_{2}=u_{3}$. Therefore $f(u)<0$ when $u<u_{1}$; $f(u)>0$ when $u_{1}<u<u_{2}$ and $u>u_{2}=u_{3}$. So the orbits of first kind which starts from the distance $u_{1}^{-1}$ and approaches to the circle of radius $u_{2}^{-1}=u_{3}^{-1}$ asymptotically by spiraling around it and the orbits of second kind starts from the distance $u_{2}^{-1}=u_{3}^{-1}$ and plunges to the singularity. 

Case $(\d)$: All the roots are real, positive and equal. Here we shall take $0<u_{1}=u_{2}=u_{3}$. Therefore $f(u)<0$ when $u<u_{1}=u_{2}=u_{3}$ and $f(u)>0$ when $u<u_{1}=u_{2}=u_{3}$. So the only orbits which can be occured in this case is the orbits of second kind. It starts from the distance $u_{1}^{-1}=u_{2}^{-1}=u_{3}^{-1}$ and plunges to
singularity.

Case $(\epsilon)$: Exactly one root is positive and the other two are complex conjugate. Here we shall take $0<u_{1}$ and $u_{2},u_{3}$ are complex. Therefore $f(u)<0$ when $u<u_{1}$ and $f(u)>0$ when $u>u_{1}$. This case is similar as Case $(\d)$, that is orbits of second kind which starts from the distance $u_{1}^{-1}$ and plunges to
singularity.

 These cases can also be described by interpreting 
\ben
V_{eff}^{2}=\b \left( 1+\frac{L^{2}}{r^{2}}\right) \left(1-\frac{2\bar{M}}{r} \right)
\label{eq:34}
\een 
as effective ``Potential Energy", which occurs from the equation (\ref{eq:9}).

\begin{figure}[h]
 \includegraphics[width=8cm, height=8cm]{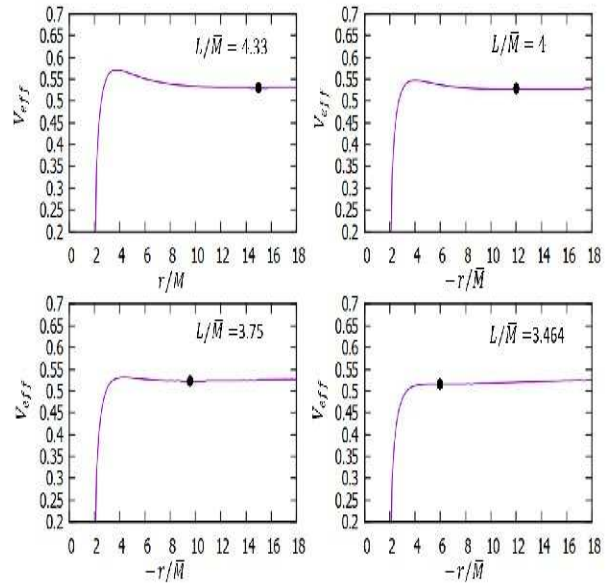}
 \caption{(Color online) The effective potentials for different time-like trajectories for $\b=0.3$ with respective to the different values of $L/\bar{M}$. Each dotted point in the figure is the minima of the corresponding potential.}\label{variation1}
\end{figure}

Fig-1 shows that the minima in the potentials for the different values of $L/\bar{M}$ correspond to the stable circular orbits while the maxima correspond to the unstable circular orbits.
Note that this figure is similar with the Schwarzschild case \cite{chandra} except the $\b$ term. 

Throughout this paper, we consider the value of $K$ i.e. dark energy density in unit of critical density \cite{planck1,planck2,planck3,planck4} is approximately $0.7$ i.e. $\beta=0.3$.

\subsubsection{Orbits of First Kind}
The orbits of the first kind are the relativistic analogues of the Keplerian orbits and to which they tend in the Newtonian limits which occurs in the cases $(\a),(\b),(\g)$ and $(\d)$. It can be parametrized by an eccentricity $e$  where $(0\le e < 1)$ and a latus rectum $l$ (some positive constant).

Let 
\ben u_{1}=\frac{1}{l}(1-e) ,
u_{2}=\frac{1}{l}(1+e) ,
u_{3}=\frac{1}{2\bar{M}}-\frac{2}{l}
\label{eq:35}
\een
where $0<u_{1}\le u_{2}\le u_{3}$. 

Now by defining $\m=\frac{\bar{M}}{l}$ and considering $u_{2}\le u_{3}$, we have 
\ben
1-6\m-2\m e\geq 0.
\label{eq:36}
\een
Also using (\ref{eq:31}), (\ref{eq:32}) and (\ref{eq:33}) we have
\ben
\frac{1}{L^{2}}=\frac{1}{\bar{M}l}\left[1-\m (3+e^{2}) \right]~;
\label{eq:37}
\een
\ben
\frac{1-\bar{E}^{2}}{L^{2}}=\frac{1}{l^{2}}(1-e^{2})(1-4\m)~;
\label{eq:38}
\een
\ben
\m < \frac{1}{3+e^{2}}~;
\label{eq:39}
\een
\ben
\m < \frac{1}{4}~;
\label{eq:40}
\een
\ben
\frac{\bar{E}^{2}}{L^{2}}=\frac{1}{\bar{M}l}\left[(1-2\m)^{2}-4\m^{2}e^{2} \right].
\label{eq:41}
\een
 Now let us take the substitution
\ben
u=\frac{1}{l}(1+e\cos\c)
\label{eq:42}
\een
Then we have at aphelion $u=\frac{(1-e)}{l}$, $\c=\pi$  and at perihelion $u=\frac{(1+e)}{l}$, $\c=0$ \cite{chandra}.

Then from the equation (\ref{eq:29}),
\ben
\left(\frac{d\c}{d\Phi}\right)^{2}=\b(1-6\m+2\m e)[1-k^{2}cos^{2}(\c/2)]
\label{eq:43}
\een
and $\P$ can be expressed in terms of the Jacobian Elliptic Integral

$$F(\Psi,k)=\int_{0}^{\Psi} \frac{1}{\sqrt{1-k^{2}\sin^{2}\n}} d\n$$
where $\Psi=\frac{1}{\pi}(\pi - \c)$,
\ben
\P=\frac{2}{\s\b (1-6\m+2\m e)^{\frac{1}{2}}} F(\frac{\pi}{2}-\frac{\c}{2},k)
\label{eq:44}
\een
where
\ben
k^{2}=\frac{4\m e}{(1-6\m+2\m e)} \le 1
\label{eq:45}
\een
and at aphelion, $\c=\pi$ and at perihelion, $\c=\frac{\pi}{2}$.The different orbits of first kind traced in Fig-2 (a), (b) and (c) with eccentricity $e=\frac{1}{2}$ , $M=\frac{3}{14}$, $\beta=0.3$, and various latus rectums $l=36,25$ and $10$ respectively on the basis of the equations (\ref{eq:42} and \ref{eq:44}) for $\bar{E}^2< 1 $.  
 
Therefore from the equations (\ref{eq:15}) and (\ref{eq:16}), the periods ( measured in co-ordinate time and in proper time ) in terms of the Newtonian period are 
\ben
\tau=\frac{\s{\b}}{2\pi} T_{Newton}(1-e^{2})^{3/2}\left[1-\m(3+e^{2})\right]^{1/2}\nonumber\\\times \int_{\c}^{\pi}d\c(1+e \cos \c)^{-2}\left[1-2\m(3+e \cos \c)\right]^{-1/2}
\label{eq:46}
\een
and 
\ben
t=\frac{1}{2\pi} T_{Newton}(1-e^{2})^{3/2}\left[(2\m-1)^{2}-4\m^{2}e^{2} \right]^{1/2} \nonumber\\\times\int_{\c}^{\pi}d\c(1+e \cos \c)^{-2}[1-2\m(1+e\cos \c)]^{-1}\nonumber\\\times[1-2\m(3+e\cos \c)]^{-1/2}\nonumber\\
\label{eq:47}
\een
where $T_{Newton}=\left(\frac{4\pi^{2}l^{3}}{(1-e^{2})^{3}GM}\right)^{1/2}$ \cite{chandra}.

Now we illustrate two special cases:

i) \textbf{Circular Orbits ($e=0$)}

In this case, we consider the two roots , $u_{1}$ and $u_{2}$ are equal (i.e., $e=0$ ). Then the orbit is a circular with the radius $l=r_{c}$ (say). From the equations (\ref{eq:37}) and  (\ref{eq:41}),
$\frac{\bar{M}}{L^{2}}=\frac{1}{r_{c}}(1-\frac{3\bar{M}}{r_{c}})$ and $\frac{\bar{M}\bar{E}^{2}}{L^{2}}=\frac{1}{r_{c}}(\frac{2\bar{M}}{r_{c}}-1)^{2}$, combining them, we get 
\ben
r_{c}=\frac{L^{2}}{2\bar{M}}\left[1\pm \sqrt{1-\frac{12\bar{M}^{2}}{L^{2}}}\right]^{1/2}.
\label{eq:48}
\een
Therefore there is no possibility of circular orbit $L/\bar{M}<2\sqrt{3}$
and also $ r_{c}=6\bar{M}$ and $\bar{E}^{2}=8/9$ for the minimum value of $L/ \bar{M}(=2 \sqrt{3})$.

Now for $L/\bar{M}>2\sqrt{3}$, the larger root of (\ref{eq:48}) locates the minimum of the  potential-energy curve $V_{eff}(r)$ which is defined by equation (\ref{eq:34}) and the smaller root locates the maximum of the potential-energy curve. Thus the circular orbit of larger radius $(6\bar{M}<r_{c}<\infty)$ will be stable and the circular orbit of smaller radius $(3\bar{M}\le r_{c}\le 6\bar{M})$ will be unstable. 

The periods of one complete revolution of these kinds  of circular orbits 
\ben
\tau_{Period}=T_{Newton} \sqrt{\b} \left(\frac{1-3\m}{1-6\m}\right)^{1/2} 
\label{eq:49} 
\een
and
\ben
t_{Period}=T_{Newton}\frac{1}{(1-6\m)^{1/2}}
\label{eq:50} 
\een
therefore,  $t_{Period}\to \infty$ when $r_{c}\to 6\bar{M}$ since $\m=\frac{\bar{M}}{l}=\frac{\bar{M}}{r_{c}}$.

ii) \textbf{Asymptotic orbits ($2\m(3+e)=1$)}

Here we consider the two roots  $u_{2}$ and $u_{3}$ are equal (i.e., $2\m(3+e)=1$), here ($0<e\le 1$). Therefore, using equation (\ref{eq:35}) the perihelion $(r_{p})$ and the aphelion $(r_{ap})$ distances are
\ben
r_{p}=2\bar{M}\frac{3+e}{1+e}
\label{eq:51}
\een
and
\ben
r_{ap}=2\bar{M}\frac{3+e}{1-e}
\label{eq:52}
\een
For these kind of orbits, the perihelion distances are restricted to the range $4\bar{M}\le r_{p}<6\bar{M}$.

From the equations (\ref{eq:37}) and (\ref{eq:41}), we have $\frac{\bar{M}^{2}}{L^{2}}=\frac{(3-e)(1+e)}{4(3+e)^{2}}$ and $1-\bar{E}^{2}=\frac{1-e^{2}}{9-e^{2}}$.
Then the equation (\ref{eq:43}) become
\ben
\left(\frac{d\c}{d\Phi}\right)^{2}=4\m e\b sin^{2}(\c/2)
\label{eq:53}
\een
or
\ben
\frac{d\c}{d\Phi}=-2(\m e\b)^{1/2}. sin(\c/2)
\label{eq:54}
\een
Here we have chosen negative value as $\Phi$ may increase when $\c$ decreases from its aphelion value $\pi$ to its perihelion value $0$.
And the solution of the equation (\ref{eq:54}) becomes
\ben
\P=-\frac{1}{\sqrt{\b \m e }}\log (\tan \frac{\c}{4})
\label{eq:55}
\een
This shows that $\P=0~when~\c=\pi$ and  $\P \to \infty$ as $\c \to 0$ and therefore  the orbits asymptotically approaches to  the perihelion by spiraling around it an infinite number of times in the counter-clockwise direction as shown in Fig-2 (d) where we consider the eccentricity $e=\frac{1}{2}$ , latus rectum $l=5$, $M=\frac{3}{14}$ and $\beta=0.3$ on the basis of the equation (\ref{eq:55}) and (\ref{eq:42}).   

iii) \textbf{The post-Newtonian approximation ($\m$ is a very small quantity)}

The Keplerian orbits of the Newtonian theory can be obtained from equation (\ref{eq:43}) considering $\m$ is very small. Then expanding the equation (\ref{eq:43}) to the first order of $\m$, we get
\ben
-d\P=\s{\b}[1+\m(3+e~cos\c)]d\c
\label{eq:56}
\een
or, in integrating,
\ben
-\P=\s{\b}(1+3\m)\c +\s{\b}\m e~sin\c +C
\label{eq:57}
\een
where $C$ is an integrating constant.

From the above equation (\ref{eq:57}), after one complete revolution during which $\c$ changes by $2\pi$, the change in $\P$ is $2\pi\s{\b}(1+3\m)$.
Therefore, the advance in the perihelion ($\D\P$) per revolution is
\ben
\D\P=6\pi\s{\b}\m=6\pi\s{\b}\frac{\bar{M}}{l}=6\pi\s{\b}\frac{\bar{M}}{a(1-e^{2})}
\label{eq:58}
\een
where $a$ denotes the semi-major axis of the Keplerian ellipse. These above results are different from the Schwarzschild case in the presence of term $\beta$.

\subsubsection{Orbits of Second Kind}
The orbits of second kind starts at a certain aphelion distance ($u_{3}^{-1}$) and plunges into singularity at $r=0$ \cite{chandra}. These kind of orbits have no Newtonian analogues. From the equation (\ref{eq:31}), we observe that all  these kind of orbits start outside the horizon since $u_{1}+u_{2}>0$ and $u_{3}<1/2\bar{M}$.

We now make the substitution
\ben
u=\left(\frac{1}{2\bar{M}}-\frac{1}{l}\right)+\left(\frac{1}{2\bar{M}}-\frac{3+e}{l}\right)\tan^{2}\x/2
\label{eq:59}
\een
in the governing equation (\ref{eq:30}), which gives
\ben
\left(\frac{d\x}{d\P}\right)^{2}=\b(1-6\m+2\m e)(1-k^{2}sin^{2}(\x/2))
\label{eq:60}
\een
or
\ben
\P=\frac{2}{\sqrt{\b(1-6\m+2\m e)}}F\left(\frac{\x}{2},k\right)
\label{eq:61}
\een
by this substitution, at aphelion , $u_{3}=\left(\frac{1}{2\bar{M}}-\frac{2}{l}\right)$, when $\x=0$ and $\P=0$; at singularity $u\to \infty$ when $\x\to \pi$ and $\P$ takes the finite value
\ben
\P_{0}=\frac{2}{\sqrt{\b(1-6\m+2\m e)}}\mathcal{K}(k)
\label{eq:62}
\een
where $\mathcal{K}(k)$ denotes the complete elliptic integral
\ben
\mathcal{K}(k)=\int_{0}^{\pi/2}\frac{1}{\sqrt{1-k^{2}\sin^{2}\n}}d\n
\label{eq:63}
\een
and $k^{2}$ as the same value of (\ref{eq:45}).The different orbits of second kind are traced in Fig-2 (a), (b) and (c) with eccentricity $e=\frac{1}{2}$ , $M=\frac{3}{14}$, $\beta=0.3$ and various latus rectum $l=36, 25$ and $10$ respectively on the basis of the equation (\ref{eq:59}) and (\ref{eq:61}) for bound orbits.
Again we consider the two cases:

i) \textbf{The case $e=0$:}

In this case, $k^{2}=0$ and 
\ben
\P=\frac{\x}{(\sqrt{\b(1-6\m)}}+\P_{c}
\label{eq:64}
\een
where $\P_{c}$ is a constant of integration and the solution for $u$ is 
\ben
u=\frac{1}{l}+\left(\frac{1}{2\bar{M}}-\frac{3}{l}\right)\sec^{2}\left[\frac{1}{2}\sqrt{\b(1-6\m)}(\P-\P_{c})\right]\nonumber\\
\label{eq:65}
\een

 Here the orbits is not a circle. The range of the aphelion distances is $3\bar{M}\le u_{3}^{-1}\le 6\bar{M}$. When $\P=\P_{c}$, this kind of orbit starts at an aphelion distance $u_{3}^{-1}$  and when $\P=\frac{\pi}{\sqrt{\b(1-6\m)}}+\P_{c}$, it arrives at the singularity ($r=0$) after circling one or more times depending on how close $\m$ to $\frac{1}{6}$ and the circle at $u_{3}^{-1}$ is the envelope of these solutions, so that circular orbit is a singular solution of the equation of motion. An example of this kind of orbit is shown in Fig-2 (e) with eccentricity $e=0$, latus rectum $l=\frac{9}{2}$ , $M=\frac{3}{14}$ and $\beta=0.3$ for bound orbits.

Now if we consider $e=0$ and $\m=\frac{1}{6}$, then all the roots of the equation $f(u)=0$ coincides and $u_{1}=u_{2}=u_{3}=\frac{1}{6\bar{M}}$ and the general solution of the governing equation (\ref{eq:30}) is
\ben
u=\frac{1}{6\bar{M}}+\frac{2}{\b\bar{M}(\P-\P{c})^{2}}
\label{eq:66}
\een
This orbit approaches to the circle at $6\bar{M}$, asymptotically, by spiraling around it an infinite number of times as shown in Fig-2 (f) with eccentricity $e=0$, $\mu=\frac{1}{6}$, $M=\frac{3}{14}$, $\beta=0.3$ and $\phi_{c}=0.1$ on the basis of the equation (\ref{eq:66}).

ii) \textbf{The case $2\m(3+e)=1$:}
For this case, since the coefficient of $tan^{2}(\x/2)$ vanishes in (\ref{eq:59}), we must consider the substitution 
\ben
u=\frac{1}{l}\left(1+e+2e \tan^{2}\frac{\x}{2}\right).
\label{eq:67}
\een 
By this substitution, we have
$u=u_{2}=u_{3}=\frac{1+e}{l}$ when $\x=0$ and $u\to \infty$ when $\x=\pi$. 
From the equation (\ref{eq:30}), we obtain
\ben
\left(\frac{d\x}{d\P}\right)^{2}=\b~4\m e~sin^{2}(\x/2)
\label{eq:68}
\een
or
\ben
\P=-\frac{1}{\sqrt{\b \m e}}~ln(\tan \frac{\x}{4})
\label{eq:77}
\een
which is exactly the same equation
(\ref{eq:55}). Therefore this orbit, $\P=0$ when $\x=\pi$ and $\P \to \infty$ when $\x\to 0$. Thus the orbit approaches the circle at $r=\frac{l}{1+e}$ , asymptotically, by spiraling around it (in the counter-clockwise direction) an infinite number of times. This behaviour is same as of the orbit of the first kind as shown in Fig-2 (d).

\subsubsection{ Orbits with imaginary eccentricities}
In this case, we are considering, the one positive real root and two imaginary roots of the equation $f(u)=0$.

Let the roots are $u_{1}=\frac{1}{2\bar{M}}-\frac{2}{l},u_{2}=\frac{1+ie}{l}$ and $u_{3}=\frac{1-ie}{l}$ where $e>0$.

Then from the equation (\ref{eq:31}), (\ref{eq:32}) and (\ref{eq:33})

\ben
\frac{1}{L^{2}}=\frac{1}{\bar{M}l}\left[1-\m(3-e^{2})\right]
\label{eq:70}
\een
\ben
\frac{1-\bar{E}^{2}}{L^{2}}=\frac{1}{l^{2}}(1+e^{2})(1-4\m)
\label{eq:71}
\een
combining, we get
\ben
\frac{\bar{E}^{2}}{L^{2}}=\frac{1}{\bar{M}l}\left[(1-2\m)^{2}+4\m^{2}e^{2}\right]
\label{eq:72}
\een
These show that $l>0$ , $\m<\frac{1}{4}$ and $1-3\m+\m e^{2}>0$ since $\bar{E}^{2}<1$. In addition, we cannot set the upper limit to $e^{2}$. 
And the range of $u$ is $\frac{1}{2\bar{M}}-\frac{2}{l}\le u<\infty$.

Let us take substitution 
\ben
u=\frac{1}{l}\left(1+e\tan \frac{\x}{2}\right)
\label{eq:73} 
\een 
in equation (\ref{eq:30}) we get
\ben
\left(\frac{d\x}{d\P}\right)^{2}=2\b\left[(6\m-1)+2\m e~sin(\x)+(6\m-1)cos(\x) \right].\nonumber\\
\label{eq:74}
\een

The range of $\x$ is $\x_{0}\le \x<\pi$ where $\tan \frac{\x_{0}}{2}=-\frac{6\m-1}{2\m e}$. 

Then solution the equation (\ref{eq:74}) is
\ben
\pm\P=\frac{1}{\sqrt{\D\b}}\int^{\p}\frac{1}{\sqrt{1-k^{2}\sin^{2}\n}} d\n
\label{eq:75}
\een
where $\D^{2}=[(6\m-1)^{2}+4\m^{2}e^{2}]$, $\sin^{2}\p=\frac{1}{\D+6\m-1}\left[\D-2\m e \sin \x-(6\m-1)\cos \x\right] $ and $k^{2}=\frac{1}{2\D}(\D+6\m-1)$.

Therefore it follows that $\sin^{2}\p=1$ at aphelion $(\x=\x_{0})$ and at singularity $(\x=\pi)$. Moreover, $\sin^{2}\p=0$ when $\x=tan^{-1}\frac{2\m e}{6\m-1}$. Thus we can say that that the range of $\p$ is $-\pi/2\le \p \le \pi/2$ when $\x_{0}\le \x\le\pi$.

Now assuming that $\P=0$ at the singularity where $\x=\pi$ and $\p=\pi/2$ the equation (\ref{eq:75}) becomes
\ben
\P=\frac{1}{\sqrt{D\b}}\left[\mathcal{K}(k)-F(\p,k)\right]
\label{eq:76}
\een
where $\mathcal{K}(k)$ denotes the complete elliptic integral and $F(\p,k)$ denotes the incomplete Jacobian integral. 

The value of $\P$ at aphelion where $\x=\x_{0}$ and $\p=-\pi/2$ is 
\ben
\P_{ap}=\frac{1}{\sqrt{D\b}}2\mathcal{K}(k).
\label{eq:77}
\een
The figures in Fig-2 (g) and (h) with latus rectum $l=3.5$, $M=\frac{3}{14}$, $\beta=0.3$ and imaginary eccentricities $e=0.01i$ and $e=0.1i$ respectively on the basis of the equation (\ref{eq:73}) and (\ref{eq:76}) for bound orbits.

\begin{figure}[h]
 \includegraphics[width=9cm, height=16cm]{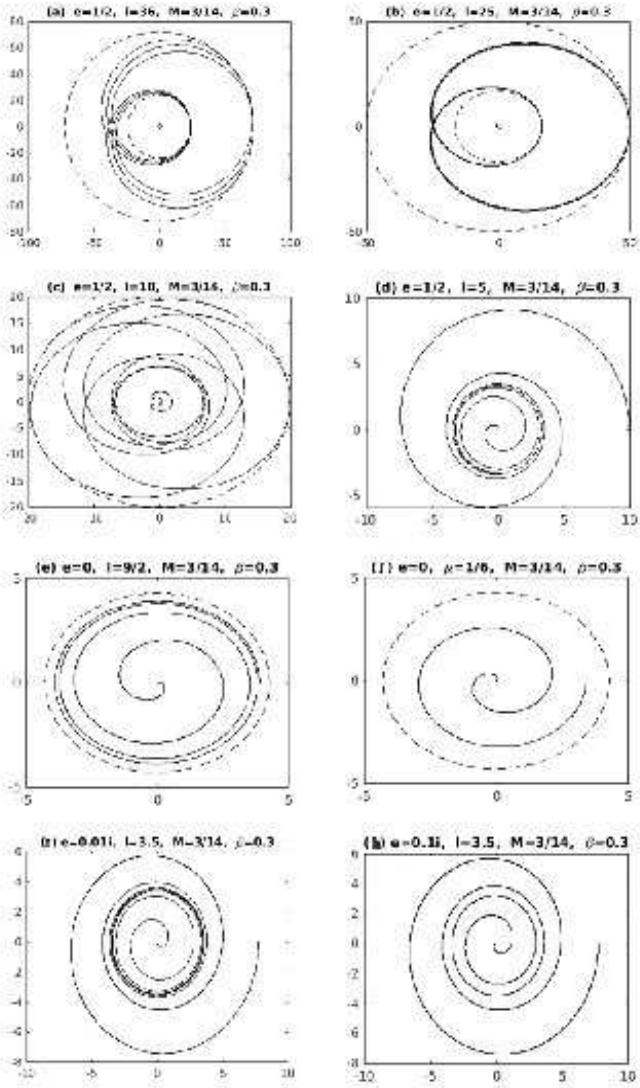}
 \caption{Timelike geodesic for the K-essence Emergent Barriola-Vilenkin spacetime for bound orbits (i.e., $\bar{E}^2<1$) and $\beta=0.3$ : (a),(b) and (c): the different kinds of orbits of First and Second Kind with real eccentricities ; (d): the orbits are of First and Second kind with eccentricity $e=\frac{1}{2}$ for the particular case $2\mu(3+e)=1$; (e): the orbits are the circular orbit for the case ($e=0$ and $\mu=\frac{1}{6}$ ) and the associated orbit of Second kind; (f): the last unstable circular orbit; (g) and (f) are the orbits with different imaginary eccentricities.}
\label{variation2}
\end{figure}

\vspace{0.1in}

\subsection{\textbf{The Unbound Orbits $(\bar{E}^{2}>1)$}}  
Here we shall restrict ourselves to unbound orbits $(\bar {E}^{2}>1)$, the governing equation (\ref{eq:13}) becomes  
\ben
g(u)=2M \left(u^{3}-\frac{1}{2\bar{M}} u^{2} +\frac{1}{L^{2}}u+\frac{\bar{E}^{2}-1}{2\bar{M} L^{2}}\right)
\label{eq:78}
\een
and 
\ben
g(u)=\left(\frac{du}{d\P}\right)^{2}
\label{eq:79}
\een
Let the roots of the cubic equation $g(u)=0$ are $u_{1},u_{2}$ and $u_{3}$ , then we have
\ben
  u_{1}+u_{2}+u_{3}=\frac{1}{2\bar{M}}
  \label{eq:80}
\een
\ben  
 u_{1}u_{2}+u_{2}u_{3}+u_{3}u_{1}=\frac{1}{L^{2}}
  \label{eq:81}
\een
\ben 
 u_{1}u_{2}u_{3}=-\frac{\bar{E}^{2}-1}{2\bar{M} L^{2}}
   \label{eq:82}
\een

Now by our assumption $(\bar {E}^{2}>1)$, the equation $g(u)=0$ has exactly one negative root and $g(u)>0$ at $u=0$. For every pair of values $\bar{E}$ and $L$, this leads to the following cases

Case $(\a)$: One root is negative and other two are positive and distinct.  Here we shall take $u_{1}<0$ and $0<u_{2}<u_{3}$. Therefore $g(u)>0$ when $0<u<u_{2}$; $g(u)<0$ when $u_{2}<u<u_{3}$ and $g(u)>0$ when $u>u_{3}$. So there are two
kinds orbits which can be traced. One is restricted in the interval $0<u<u_{2}$. That is, these kind of orbits are analogues of the hyperbolic orbits of the Newtonian Theory. We will call these kind of orbits is the orbit of first kind. And
the other is $u > u_{3}$ which we will call the orbits of second kind.
	 
Case $(\b)$: One root is negative and other two are positive and equal.  Here we shall take $u_{1}<0$ and $0<u_{2}=u_{3}$. Therefore $g(u)>0$ when $0<u<u_{2}=u_{3}$ and $g(u)>0$ when $u>u_{2}=u_{3}$. The orbits of first kind arrives from infinity
and approaches to the circle of radius $u_{2}^{-1}=u_{3}^{-1}$ asymptotically by spiraling
around it. And the orbits of second kind starts from the distance $u_{3}^{-1}$ and
plunges to singularity.

Case $(\g)$: One root is negative and other two are complex conjugate. Here we shall take $u_{1}<0$ and $u_{2},u_{3}$ are complex-conjugate. Therefore $g(u)>0$ when $u>0$. This implies the orbits of second kind can occur but there is no
possibilities of the occurrence of the orbit of first kind. The orbits of second kind
starts from infinity and plunges to the singularity.
  
\subsubsection{Orbits of First Kind and Second Kind}
The orbits of the first and second kind occurs in the cases $(\a)$ and $(\b)$.
Here we are not bothered about the  calculation of the unbound orbits of the second kind as they are same as the bound orbits of the second kind with the only difference is $e\ge 1$.    
  When $u_{2}=u_{3}$, the two kinds of orbits coalesce as they approach a common circle from opposite sides by spiraling round it an infinite number of times asymptotically.

Let the roots are 
\ben u_{1}=-\frac{1}{l}(e-1) ,
u_{2}=\frac{1}{l}(1+e) ,
u_{3}=\frac{1}{2\bar{M}}-\frac{2}{l}
\label{eq:83}
\een
where $u_{1}<0$ and $0< u_{2}\le u_{3}$ and $e\ge 1$. 

Again by defining $\m=\frac{\bar{M}}{l}$ , we have 
\ben
1-6\m-2\m e\ge 0
\label{eq:84}
\een
and using (\ref{eq:80}), (\ref{eq:81}) and (\ref{eq:82}), we have
\ben
\frac{1}{L^{2}}=\frac{1}{\bar{M}l}\left[1-\m (3+e^{2}) \right]
\label{eq:85}
\een
\ben
\frac{\bar{E}^{2}-1}{L^{2}}=\frac{1}{l^{2}}(e^{2}-1)(1-4\m)
\label{eq:86}
\een
since $L^{2}>1$ and $\bar{E}^{2}-1 \ge 0$, we have
\ben
\m < \frac{1}{3+e^{2}}
\label{eq:87}
\een
\ben
\m \le \frac{1}{4}
\label{eq:88}
\een
Now the inequalities (\ref{eq:84}) and (\ref{eq:87}) ensure that the range of $e$ is $1\le e <3$. 
 Now let us take the substitution
\ben
u=\frac{1}{l}(1+e\cos\c)
\label{eq:89}
\een
then the allowed range of $\c$ is $0\le \c < \c_{\infty}$ where $\c_{\infty}=\cos^{-1}(-\frac{1}{e})$.
Therefore the solution of the equation (\ref{eq:43}) is 
\ben
\P=\frac{2}{\sqrt{\b(1-6\m+ 2\m e)}}\left[\mathcal{K}(k)-F(\pi/2-\c/2,k)\right]\nonumber\\
\label{eq:90}
\een
where $\p=\pi/2$ when $\c=0$ and $\p=\frac{1}{2}\cos^{-1}(\frac{1}{e})=\p_{\infty}$(say) when $\c=\c_{\infty}$ and $k^{2}=\frac{4\m e}{1-6\m+2\m e}$.
From the equation (\ref{eq:90}), it can be explained that $\P=0$ when $\c=0$ and 
\ben
\P=\frac{2}{\sqrt{\b(1-6\m+ 2\m e)}}\left[\mathcal{K}(k)-F(\p_{\infty},k)\right]=\P_{\infty}(say)\nonumber\\
\label{eq:91}
\een
when $\p=\p_{\infty}$. 

Therefore these kind of orbit goes to infinity, asymptotically, along the direction $\P_{\infty}$. In Fig-3 (a), (b) and (c) , the orbits of first kind and second kind are traced with eccentricity $e=\frac{3}{2}$, $M=\frac{3}{14}$, $\beta=0.3$ and various latus rectums $l=15,\frac{17}{2}$ and $\frac{14}{2}$ respectively on the basis of the equations (\ref{eq:90}) and (\ref{eq:89}) for First kind and equations (\ref{eq:59}) and (\ref{eq:61}) for Second kind.  

In this context, if we consider the special case when the two positive roots of $g(u)=0$ are equal i.e., $2\m(3+e)=1$, 

The equations (\ref{eq:85}) and (\ref{eq:86}) becomes 
\ben 
\frac{L^{2}}{\bar{M}^{2}}=\frac{4(3+e)^{2}}{(3-e)(1+e)}
\label{eq:92}
\een  
and
\ben 
\bar{E}^{2}-1=\frac{e^{2}-1}{9-e^{2}}.
\label{eq:93}
\een
In this case the corresponding perihelion distance is 
$$r_{p}=\frac{2\bar{M}(3+e)}{1+e}$$
and the range of the perihelion distances is $3\bar{M}< r_{p}\le 4\bar{M}$.

And the solution the equation (\ref{eq:79}) is
\ben
\P=-\frac{1}{\sqrt{\b\m e}}\log (\tan \c/4)
\label{eq:94}
\een
Thus we can say that the unbound orbit approaches the circle at $r_{p}$, asymptotically, by spiraling around it an infinite number of times.  

\subsubsection{The orbits with imaginary eccentricities}
In this case, we are considering, the one negative real root and two imaginary roots of the equation $g(u)=0$.

Let the roots are $u_{1}=\frac{1}{2\bar{M}}-\frac{2}{l},u_{2}=\frac{1+ie}{l}$ and $u_{3}=\frac{1-ie}{l}$ where $e>0$ and $u_{3}<0$ .

Then from the equation (\ref{eq:80}), (\ref{eq:81}) and (\ref{eq:82})

\ben
\frac{1}{L^{2}}=\frac{1}{l\bar{M}}\left[1-\m(3-e^{2})\right]
\label{eq:95}
\een
\ben
\frac{\bar{E}^{2}-1}{L^{2}}=\frac{1}{l^{2}}(1+e^{2})(4\m-1)
\label{eq:96}
\een

These shows that $\m\ge \frac{1}{4}$ and $1-3\m+\m e^{2}>0$. The values of $\m$ in the range $\frac{1}{3}\le \m \le \frac{1}{4}$, there is no restriction on $e^{2}$ but when $\m>\frac{1}{3}$, the necessary restriction for $e^{2}$ is $e^{2}>(3-\frac{1}{\m})$. 

Let us take the same substitution as (\ref{eq:73}), we have $u=0$ when $\x=2\tan^{-1}(-\frac{1}{e})=\x_{\infty}$ (say), $u<0$ when $\x<\x_{\infty}$ and $u\to \infty$ when $\x=\pi$. Thus the range of $\x$ is $\x_{\infty}<\x \le \pi$. 
Apart from these changes from bound orbit case, the solution of the equation (\ref{eq:79}) is
\ben
\P=\frac{1}{\sqrt{\D\b}}\left[\mathcal{K}(k)-F(\p,k)\right]
\label{eq:97}
\een
where $\D^{2}=(6\m-1)^{2}+4\m^{2}e^{2}$, $\sin^{2}\p=\frac{1}{\D+6\m-1}\left[\D-2\m e \sin \x-(6\m-1)\cos \x\right] $ and $k^{2}=\frac{1}{2\D}(\D+6\m-1)$.

Therefore, $\P=0$ at $\x=\pi$ i.e. at $\p=\pi/2$ and the orbit goes to infinity when $\x=\x_{\infty}$ i.e., at $\p=\p_{\infty}$ where 
$\sin \p_{\infty}=\frac{1}{\D+6\m-1}\left[\D+6\m-1-2e^{2}\frac{4\m-1}{e^{2}+1}\right]$.
Lastly in Fig-3 (d) and (e), we have traced the orbits with latus rectum $l=3.3$, $M=0.3$, $\beta=0.3$ and with imaginary eccentricities $e=0.001i$ and $e=0.1i$ respectively on the basis of the equations (\ref{eq:97}) and (\ref{eq:73}).

\begin{figure}[h]
 \includegraphics[width=9cm, height=14cm]{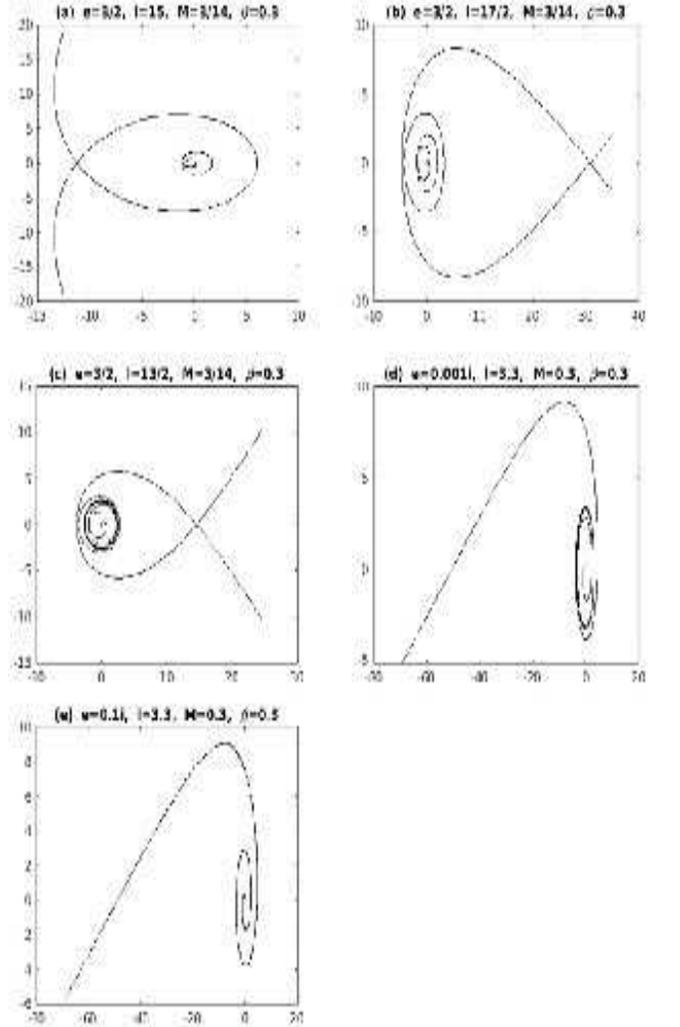}
 \caption{Timelike geodesic for the K-essence Emergent Barriola-Vilenkin spacetime for unbound orbits ( i.e., $\bar{E}^2>1$) and $\beta=0.3$ : (a),(b) and (c) are the different kinds of orbits of First and Second Kind with real eccentricities;(d) and (e) are the orbits with different imaginary eccentricities.}
\label{variation2}
\end{figure}

\vspace{5in}

\section{Conclusion}
We report the exciting features of the time-like geodesic structure in the presence of dark energy in an emergent gravity scenario for this Barriola-Vilenkin type metric based on S. Chandrasekhar. In the presence of the dark energy density $(K)$ of the BV type spacetime, for radial geodesic, the time taken for a particle cross the horizon is greater than that of Schwarzschild spacetime since $K$ has values $0<K<1$. 
	
On exploring the existence of the bound orbits for $E<\sqrt{\beta}$ and the unbound orbits for $E>\sqrt{\beta}$,
we have completely evaluated the periods which are measured in coordinate time and in proper time for the non-radial bound orbits of the first kind in the case of time-like geodesics. 

For $E<\sqrt{\beta}$, no circular orbit is possible for $L/M<2\sqrt{3\beta}$. The range of the perihelion distances for the asymptotic orbits of first kind is  $4M/\sqrt{\beta}\le r_{p}\le 6M/\sqrt{\beta}$, while that of the aphelion distances for the orbits of second kind with zero eccentricity is $3M/\sqrt{\beta}\le r_{ap}\le 6M/\sqrt{\beta}$.
In the condition of $E>\sqrt{\beta}$, the equality of the two positive roots of the equation $g(u)=0$, for the unbound orbits, makes the range of the perihelion distances $3M/\sqrt{\beta}<r_{p}\le 4M/\sqrt{\beta}$. All the ranges noted here are much higher than those in the Schwarzschild spacetime since $0 < \b < 1$. Albeit, the bound orbits of imaginary eccentricity are starting from a finite aphelion distance and the unbound orbits of imaginary eccentricity are starting from infinity, they fall into the singularity. Finally, we conclude that the basic structure of the trajectories of both the bound and unbound orbits in the {\bf k-}essence emergent BV -type spacetime appears to be same, as is seen in the case of Schwarzschild spacetime. However, the allowed ranges of the aphelion and perihelion distances for the bound and unbound orbits are much higher in the {\bf k-}essence emergent BV -type spacetime in the condition, $0 < \b < 1$.

\vspace{0.2in}
{\bf Acknowledgement:} The authors would like to thank the referees for illuminating suggestions to improve the manuscript.

\end{document}